\documentclass[pdflatex,sn-nature]{sn-jnl}

\usepackage{graphicx}
\usepackage{amsmath,amssymb,amsfonts}
\usepackage{upgreek}
\usepackage[version=4]{mhchem}
\usepackage{mathtools}
\usepackage{siunitx}
\usepackage{hyperref}
\usepackage{xcolor}
\usepackage{booktabs}

\usepackage{fix-cm}

\begin{document}

\title[PHIPNOE]{Robust nuclear hyperpolarization of small molecules through intermolecular transfer of parahydrogen-derived polarization}

\author[1]{\fnm{Bogdan A.} \sur{Rodin}}\equalcont{These authors contributed equally to this work.}
\author[1]{\fnm{Anna} \sur{Parker}}\equalcont{These authors contributed equally to this work.}
\author[1,2]{\fnm{Laurynas} \sur{Dagys}}
\author[1]{\fnm{Vitaly} \sur{Kozinenko}}
\author[3,4]{\fnm{Martin} \sur{Korzeczek}}
\author[3,4]{\fnm{Martin B.} \sur{Plenio}}
\author[1,5]{\fnm{Salvatore} \sur{Mamone}}
\author[1]{\fnm{Rokas} \sur{Šakalys}}
\author[6]{\fnm{Federico} \sur{De Biasi}}
\author[6]{\fnm{Ran} \sur{Wei}}
\author[6]{\fnm{Pinelopi} \sur{Moutzouri}}
\author[6]{\fnm{Lyndon} \sur{Emsley}}
\author[1]{\fnm{Stephan} \sur{Knecht}}
\author*[1,7]{\fnm{James} \sur{Eills}}\email{j.eills@fz-juelich.de}
\author*[1]{\fnm{Ilai} \sur{Schwartz}}\email{ilai@nvision-imaging.com}

\affil[1]{\orgname{NVision Imaging Technologies GmbH}, \orgaddress{\city{Ulm}, \postcode{89081}, \country{Germany}}}
\affil[2]{\orgdiv{Institute of Chemical Physics}, \orgname{Vilnius University}, \orgaddress{\street{Saulėtekio Av. 3}, \city{Vilnius}, \postcode{LT-10257}, \country{Lithuania}}}
\affil[3]{\orgdiv{Institut für Theoretische Physik}, \orgname{Universität Ulm}, \orgaddress{\street{Albert-Einstein Allee 11}, \city{Ulm}, \postcode{D-89081}, \country{Germany}}}
\affil[4]{\orgname{Center for Integrated Quantum Science and Technology (IQST)}, \orgaddress{\city{Ulm}, \postcode{89081}, \country{Germany}}}
\affil[5]{\orgdiv{Dipartimento di Medicina, Scienze della Vita e dell'Ambiente}, \orgname{Università degli studi dell'Aquila}, \orgaddress{\city{L'Aquila}, \postcode{67100}, \country{Italy}}}
\affil[6]{\orgdiv{Institut des Sciences et Ingénierie Chimiques}, \orgname{École Polytechnique Fédérale de Lausanne}, \orgaddress{\city{Lausanne}, \postcode{CH-1015}, \country{Switzerland}}}
\affil[7]{\orgdiv{Institute for Biological Information Processing (IBI-7: Structural Biochemistry)}, \orgname{Forschungszentrum Jülich}, \orgaddress{\city{Jülich}, \postcode{52428}, \country{Germany}}}

\abstract{Nuclear magnetic resonance (NMR) spectroscopy has transformed our understanding of molecular structure and dynamics in chemistry, biology and medicine. The recent advent of hyperpolarization techniques~\cite{Eills2023}---which can enhance NMR signals by several orders of magnitude relative to thermally polarized samples---has enabled applications traditionally out of reach due to the inherently low sensitivity of NMR techniques. However, a high barrier to entry remains, as most hyperpolarization approaches either require complex instrumentation~\cite{Ardenkjaer2003,Walker1997} or are applicable only to a relatively small set of molecules. Here we introduce PHIPNOE, a platform that directly addresses both limitations. PHIPNOE is based on parahydrogen-induced polarization (PHIP)~\cite{Bowers1987, Natterer1997}, which is well-established as a scalable route to hyperpolarization requiring minimal instrumentation, but has been mostly restricted to molecules that undergo specific chemical reactions. We overcome this barrier by tailoring PHIP to create highly polarized, highly concentrated solutions of one specific molecule, which acts as an intermediate source of polarization. This `source molecule' then distributes polarization to a broad range of target molecules mixed into the solution, via the spin polarization-induced nuclear Overhauser effect (SPINOE)~\cite{Navon1996}. 
We investigate chemical influences on PHIPNOE, and develop a predictive model to estimate enhancement based on molecular mass and T$_1$ relaxation times. 
A complete run from PHIP hyperpolarization to PHIPNOE polarization transfer and signal detection takes less than one minute, the approach does not require any modifications to the NMR spectrometer, and enhancements are repeatable across molecular classes. PHIPNOE thus enables applications including single-shot multidimensional NMR, real-time monitoring of dynamic processes, and---with 300-fold signal amplification demonstrated on a benchtop spectrometer---practical low-field NMR, where we show enhanced sensitivity in detecting per- and polyfluoroalkyl substances (PFAS).}


\maketitle

\section*{Main}\label{sec1}

Nuclear magnetic resonance (NMR) spectroscopy is a central analytical technique in chemistry, biology and medicine. Because NMR signals are inherently weak, technological developments have focused on maximizing the strength of the magnets used in NMR spectrometers, for improving both resolution and sensitivity. This need for expensive and sophisticated instrumentation has meant that, while NMR is a vital tool for research, it is not a widely accessible tool for most industrial or public-sector applications. Portable, affordable and simple-to-use benchtop NMR spectrometers have recently grown rapidly in industrial, research and educational settings~\cite{galvan_successful_2023, castaing-cordier_chapter_2021}, but these systems come at the cost of decreased sensitivity and resolution. To address such sensitivity issues, methods have been developed in which a non-equilibrium state for the nuclear-spin system is created through interactions with electron or photon degrees of freedom or by exploiting spin statistics. Such `hyperpolarization' techniques can enhance NMR signals by up to four to five orders of magnitude, transforming previously unfeasible experiments into routine measurements~\cite{Eills2023}.

However, significant barriers limit the use of hyperpolarized NMR. Approaches based on dissolution dynamic nuclear polarization (dDNP) for solution NMR are broadly applicable, but require complex cryogenic and microwave instrumentation together with long experiment setup times~\cite{denysenkov_liquid_2010,Ardenkjaer2003,Jannin2019}. In representative ligand studies at 9.4 T, reported $^1$H enhancements are in the range of several hundred~\cite{kimApplicationsDissolutionDNPNMR2019,wangDeterminationProteinLigand2020}. By contrast, experimentally simpler methods are typically limited in substrate scope: spin-exchange optical pumping (SEOP)~\cite{Barskiy2016} applies to atoms or paramagnetic centres with suitable electronic transitions, whereas parahydrogen-induced polarization (PHIP)~\cite{Hoevener2018,Reineri2021} is generally tied to compounds undergoing hydrogenation reactions.

Several strategies have been developed to broaden the scope of hyperpolarized NMR by transferring polarization from a highly polarized source molecule to a target. One branch is represented by exchange-mediated relay methods such as SABRE-relay and PHIP-relay, which can produce very large $^1$H enhancements (several hundred-fold) for chemically compatible substrates, but with strong dependence on exchangeable protons and substrate chemistry~\cite{raynerRelayedHyperpolarizationParahydrogen2019,themParahydrogenInducedPolarizationRelayed2021}. Another branch relies on direct intermolecular polarization transfer via cross-relaxation. Early proof-of-principle experiments used sources such as laser-polarized xenon~\cite{Navon1996, pietrassSurfaceSelective1H1998} and $^{13}$C-hyperpolarized molecules~\cite{Marco-Rius2014}, but the enhancements were limited due to low gyromagnetic ratio and low concentrations of the source species. More recently, approaches using $^1$H as the source polarization nucleus have been implemented, achieving larger enhancements on both high field and benchtop systems~\cite{Korchak2021,Salnikov2022,Eichhorn2022,Parker2023}. Of these, two PHIP-based approaches, PRINOE~\cite{Korchak2021} and the related PAINTER~\cite{Salnikov2022,trofimovIntermolecularNuclearSpin2026}, provide signal enhancements on the order of ten-fold, but rely on RASER-specific conditions that cannot be easily translated to benchtop instruments.

Despite these advances, transferred-hyperpolarization approaches have largely remained proof-of-concept demonstrations rather than practical tools. Without a framework for predicting which molecules can be enhanced and by what magnitude, and without a protocol that ensures repeatable use at practical concentrations, experiments cannot be designed with confidence. This gap persists despite the increasing availability of hyperpolarizer units in NMR labs worldwide~\cite{coffey_open-source_2016}.

Here we introduce a framework for PHIP-based nuclear Overhauser effect (PHIPNOE) polarization transfer that closes this gap and enables repeatable hyperpolarization of a broad range of small molecules with predictable enhancements. Leveraging recent advances in maintaining PHIP efficiency at high concentrations~\cite{Dagys2024}, we designed optimized pulse sequences for polarization generation and transfer, hyperpolarizing molecules across various classes and achieving enhancements up to 30-fold at \SI{400}{\mega\hertz} and up to 300-fold at \SI{80}{\mega\hertz}. In parallel, we developed a simple model based on T$_1$ relaxation times and molecular mass that reliably predicts enhancement levels. Finally, we use heteronuclear PHIPNOE to improve the sensitivity of \textsuperscript{19}F NMR detection of per- and polyfluoroalkyl substances (PFAS).

Our results establish PHIPNOE as a broadly applicable approach to hyperpolarization that overcomes constraints limiting previous methods. In comparison to approaches based on DNP, PHIPNOE benefits from the well-established advantages of PHIP-based methods, including operation at room temperature without microwave irradiation or cryogenics, preparation times of only minutes, and compatibility with existing standard NMR hardware. Combined with the predictability afforded by our framework, PHIPNOE makes hyperpolarization practical for standard NMR laboratories: the complete procedure takes only one minute, requires no modifications to the NMR spectrometer, and achieves enhancements up to 300-fold on benchtop instruments. These capabilities enable applications currently limited by low polarization, including single-shot multidimensional NMR, real-time reaction monitoring, and benchtop analysis of dilute samples.

\subsection*{The PHIPNOE hyperpolarization platform}

\begin{figure}[h]
    \centering
    \includegraphics[width=\textwidth]{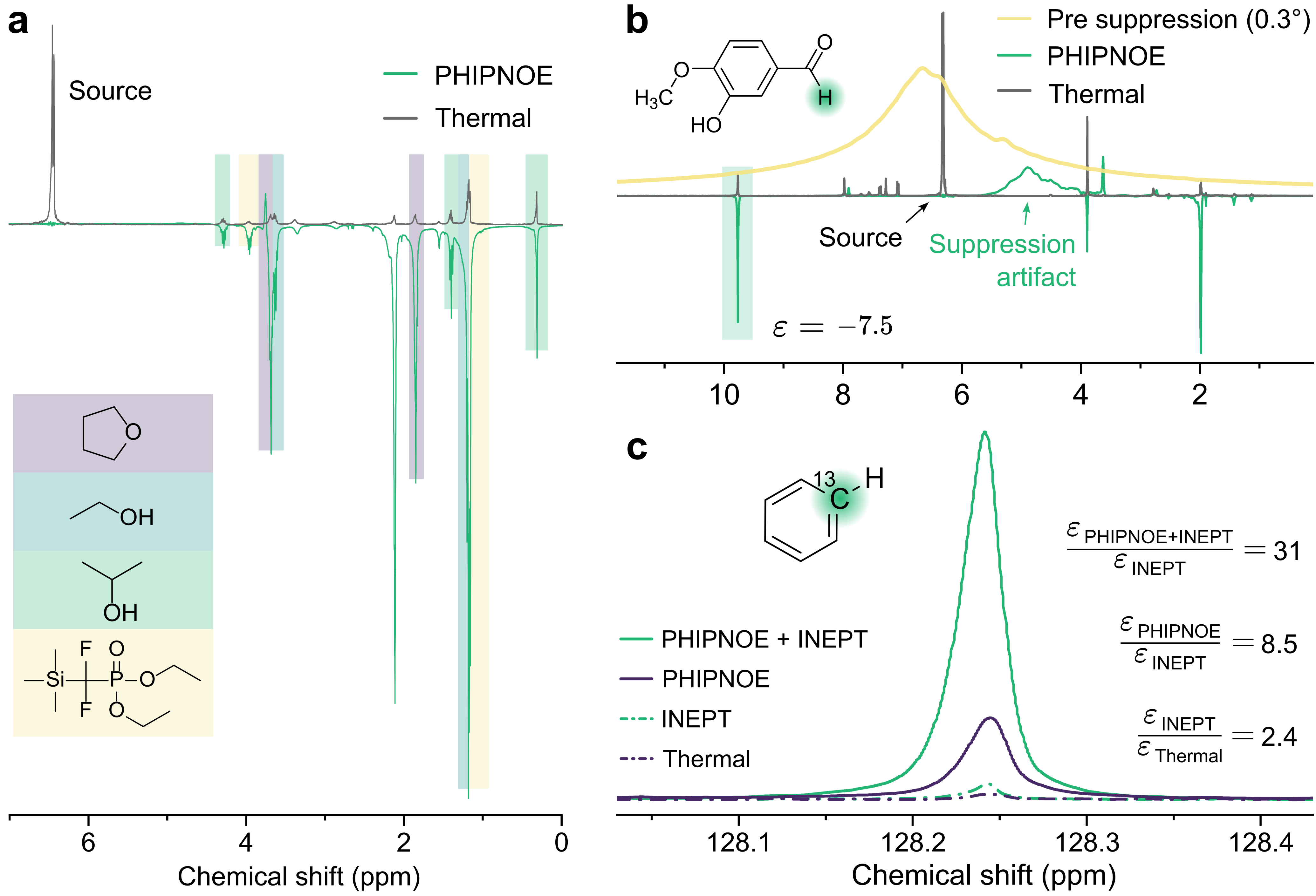}
    \caption{\textbf{Versatility of the PHIPNOE hyperpolarization platform.} \textbf{a}, Simultaneous hyperpolarization of a mixture containing THF, ethanol, isopropanol and diethyl [difluoro(trimethylsilyl)methyl]phosphonate. The spectrum was recorded eight seconds after polarization transfer. \textbf{b}, Single-shot spectrum of isovanillin, recorded at the optimal time derived from its build-up kinetics (similar to Fig.~\ref{fig:RD_and_kinetics}). For both \textbf{a} and \textbf{b}, strong solvent suppression was used to eliminate the source signal (discussed in the main text). \textbf{c}, $^{13}$C polarization in benzene. A comparison is shown between $^{13}$C NMR signals: (1) at thermal equilibrium, (2) following $^1$H$\rightarrow ^{13}$C INEPT transfer, (3) enhanced directly via PHIPNOE, and (4) enhanced by a combination of $^1$H PHIPNOE and $^1$H$\rightarrow ^{13}$C INEPT. The ratios of $^{13}$C signal enhancements ($\varepsilon$) in these experiments are shown to the right.}\label{fig:teaser}
\end{figure}

In order to harness the advantages of the PHIPNOE approach for the general hyperpolarization of target molecules, we developed an experimental platform that combines efficient parahydrogen-based polarization generation with controlled intermolecular transfer. Our approach builds on the principle that hyperpolarized molecules in solution can transfer their polarization to surrounding species through dipolar cross-relaxation, provided the experimental conditions are optimized for the process. The versatility of PHIPNOE is showcased in Fig.~\ref{fig:teaser}, which shows hyperpolarization of different targets: (a) four molecules (THF, ethanol, isopropanol and diethyl [difluoro(trimethylsilyl)methyl]phosphonate) present in solution for the entire PHIPNOE process; (b) isovanillin after injection of a small quantity into the polarized source solution; and (c) $^{13}$C in benzene. A comparison is shown between three ways of enhancing the $^{13}$C signal: $^1$H$\rightarrow ^{13}$C INEPT (insensitive nuclei enhanced by polarization transfer) at thermal equilibrium, direct PHIPNOE (dipolar cross-relaxation between $^1$H spins in the source species and the $^{13}$C spin in benzene), and PHIPNOE to polarize the benzene $^1$H spins followed by $^1$H$\rightarrow ^{13}$C INEPT.

\begin{figure}[h]
    \centering
    \includegraphics[width=\textwidth]{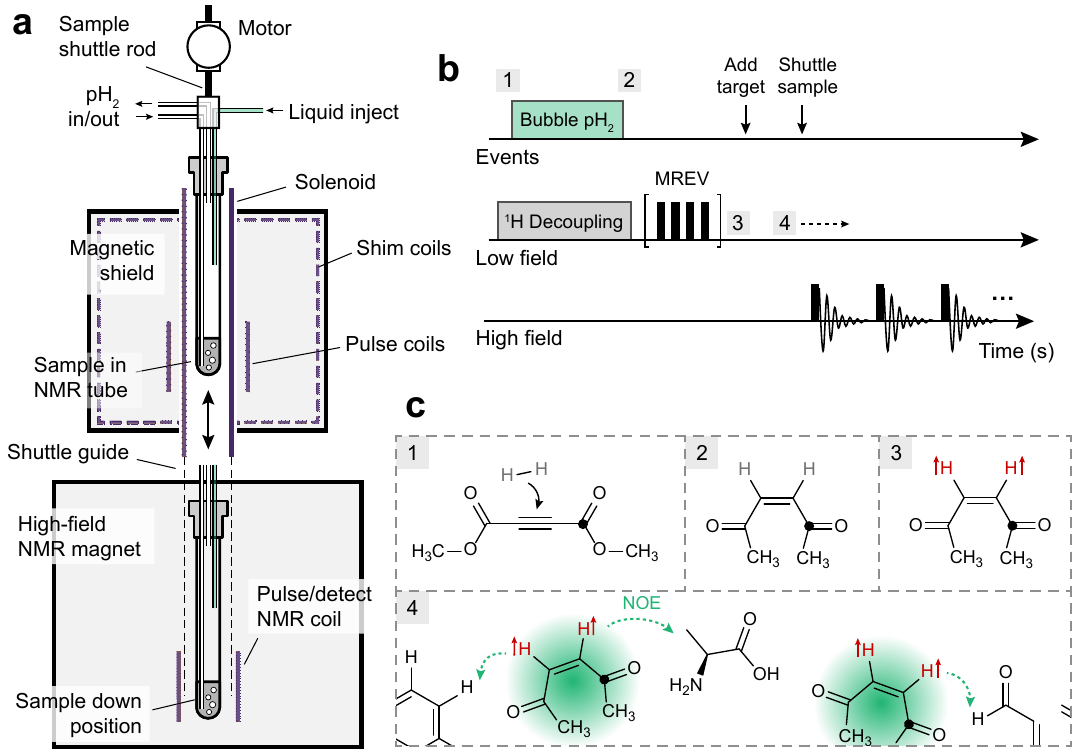}
    \caption{\textbf{PHIPNOE experimental workflow: from parahydrogen to target hyperpolarization.} \textbf{a}, The hydrogenation and application of the polarization transfer pulse sequence were performed in the magnetic shield. The sample is then transferred into the high-field region using a mechanical shuttle system. The majority of target molecules were added using an additional outlet for liquid injection. For more details, see `Experimental protocols' in the Methods section. \textbf{b}, The main experimental protocol for extracting the largest enhancement factor for the target molecules. We employ a pulse sequence that enables polarization transfer while simultaneously suppressing distant dipolar fields, conceptually similar to the approach described in ref.~\cite{Dagys2024}. A small excitation angle is used to track the signal over time to obtain the maximum enhancement. The number labels refer to the steps shown in panel \textbf{c}. \textbf{c}, Molecular level scheme, where parahydrogen is attached to dimethyl acetylenedicarboxylate (1), forming dimethyl maleate (2). Polarization transfer occurs where singlet order is converted to $^1$H magnetization (3). This large magnetization is then transferred to other molecules in solution by intermolecular NOE (4).}\label{fig:setup}
\end{figure}

We chose [1-$^{13}$C]dimethyl maleate-d$_6$ (DMM) as our polarization source molecule, generated by the pairwise addition of parahydrogen to [1-$^{13}$C]dimethyl acetylene dicarboxylate-d$_6$ (DMAD). This system offers high hydrogenation and polarization transfer efficiency, good solubility in various solvents, favourable relaxation properties for NOE transfer, and safe handling. The experimental workflow proceeds in three stages: polarization generation at low field, rapid transfer to high field, and controlled mixing with target molecules while monitoring the polarization dynamics (Fig.~\ref{fig:setup}).

\begin{figure}[h]
    \centering
    \includegraphics[width=0.9\textwidth]{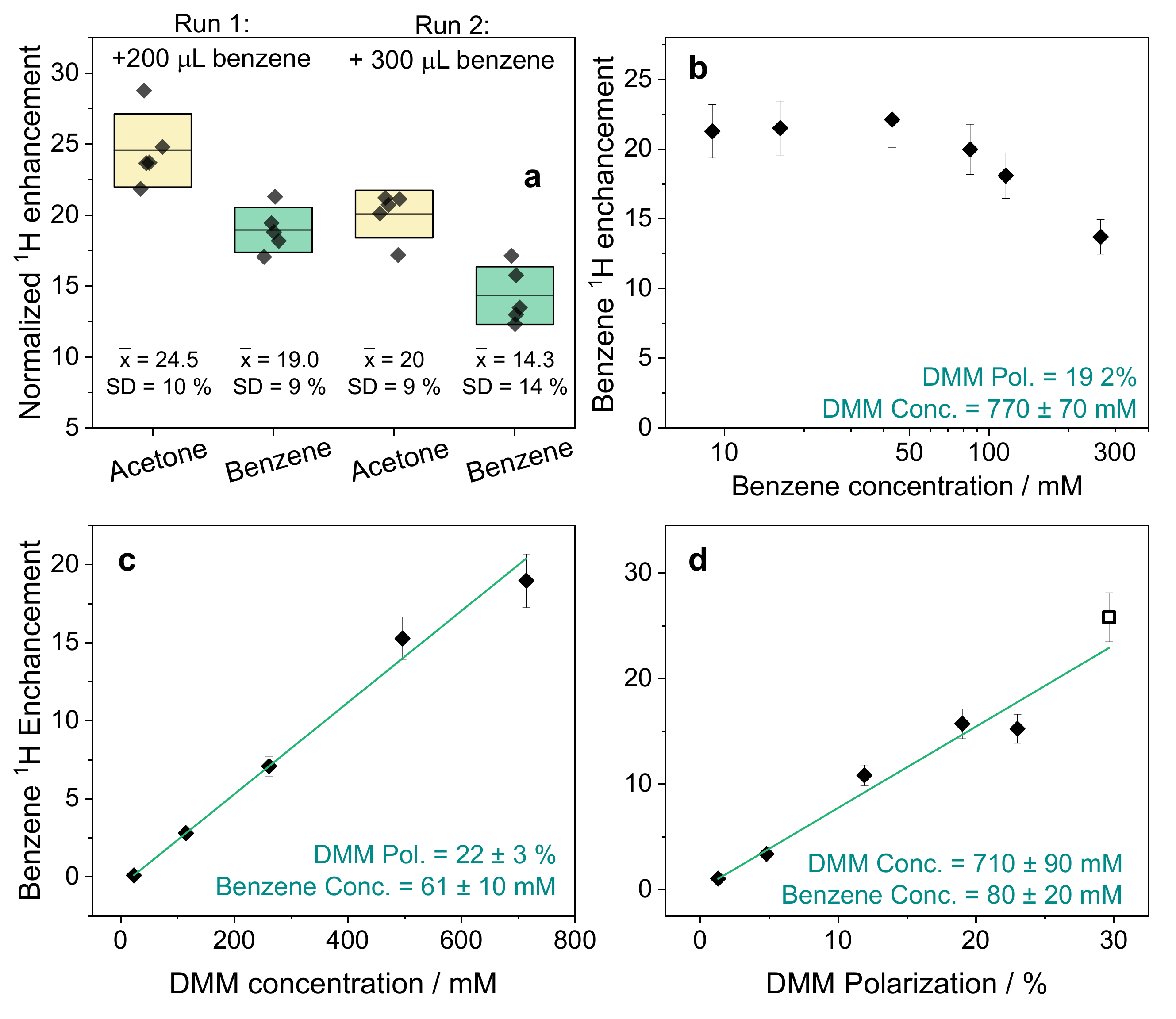}
    \caption{\textbf{Repeatability of PHIPNOE enhancements and dependence on experimental factors.} \textbf{a}, $^1$H signal enhancement for the solvent acetone C(CD$_3$)(CD$_2$H) isotopomer, and for benzene molecules added to the source-molecule solution via a capillary. The centres of the box correspond to the mean and the edges to the standard deviation. Run \#1 and \#2 correspond to the addition of \SI{200}{\micro\L} and \SI{300}{\micro\L} of \SI{400}{mM} benzene solution, respectively, to the hyperpolarized DMM solution. \textbf{b--d,} Benzene $^1$H enhancement as a function of benzene concentration (\textbf{b}), DMM concentration (\textbf{c}) and DMM polarization (\textbf{d}). In (\textbf{d}), the empty point corresponds to the run without dilution and with the narrow glass capillary to limit evaporation during bubbling.}\label{fig:enhancements}
\end{figure}

In a typical PHIPNOE experiment, we dissolved \SI{1}{M} DMAD and \SI{10}{mM} rhodium catalyst in acetone-d$_6$. The solution (\SI{400}{\micro\L}) was placed in a pressurizable NMR tube held in a magnetic shield, with a \SI{100}{\micro\tesla} magnetic field provided by a solenoid coil. Parahydrogen was bubbled through the solution (\SI{10}{bar}, \SI{30}{s}), with concurrent $^1$H decoupling pulses to mitigate singlet--triplet mixing during hydrogenation~\cite{kating_nuclear_1993}. For converting the proton singlet order to observable magnetization, we developed a pulse sequence that enables polarization transfer while simultaneously suppressing the interaction between the $^1$H nuclear spins and their self-induced distant dipolar field~\cite{korzeczek_phip_2025}. Such suppression is crucial for preserving polarization at high concentration levels~\cite{Dagys2024}. Following this step, a solution of the target molecules was injected, and the mixed sample was rapidly shuttled to a 9.4-T NMR magnet. This procedure yielded on average 740$\pm$60\,mM DMM at 19$\pm$2\% polarization after injection of the target molecule solution (see the Methods section for details). Given that each DMM molecule has two polarized $^1$H spins, the molar polarization\cite{knecht_rapid_2021} of the source molecules was 280$\pm$20 mM, which serves as a benchmark figure of merit. The entire PHIPNOE procedure takes only 45 seconds, and when combined with automated sample handling, it can be immediately repeated for the next sample. This systematic procedure was applied for the experiments shown in Fig.~\ref{fig:enhancements} (Run~\#1 in panel a, and panels b--d). The demonstrator and benchtop experiments (Figs~\ref{fig:teaser} and~\ref{fig:f80}) were performed without injections and are described in subsequent sections.

We characterized the repeatability and parameter dependencies of the technique using the following procedure: after DMM polarization, \SI{200}{\micro\L} of target molecule solution (360~mM benzene in acetone-d$_6$) was injected under pressure (\SI{5}{bar} of N$_2$ pressure for \SI{8}{s}), followed by sample shuttling to a 9.4-T magnet and signal acquisition every \SI{5}{s}. We used benzene as a reference target molecule due to the simplicity of its spectrum and large PHIPNOE enhancements. The PHIPNOE process contains a number of steps that may introduce variability, such as the H$_2$ bubbling efficiency, polarization transfer sequence, and target solution mixing efficiency. Fig.~\ref{fig:enhancements}a shows the results of repeatability tests in which we performed the experiment from start to finish with a fresh sample each time, and measured the PHIPNOE enhancements on two molecules: the acetone solvent which is in solution for the entire procedure, and benzene which was injected after the source-molecule polarization step. On both molecules we observed consistent enhancement factors, with standard deviations below 15\% across multiple runs ($N=5$ runs for each box). In Run \#1 we injected \SI{200}{\micro\L} of the target molecule solution, and in Run \#2 we injected \SI{300}{\micro\L}. Both variants showed similar polarization repeatability. We note that in Run \#2 the signal enhancement on benzene is 25\% lower, which is expected and corresponds (within experimental error) to the 33\% lower DMM:benzene concentration ratio. We attribute the observed variability to fluctuations in bubbling and injection efficiency, as well as potential fitting errors.

Having established the repeatability of the technique, we next examined how enhancement depends on experimental parameters. First, at concentrations below 50 mM, the enhancement on the $^1$H nuclear spins of the target molecule was essentially independent of target concentration. At a high target concentration of \SI{250}{mM}, the enhancement decreased by $\approx$ 30\% (see Fig.~\ref{fig:enhancements}b). This is an encouraging result, as low concentrations of target molecules are typically preferred for applications of hyperpolarized NMR. In addition, the final enhancement on the target molecule depended linearly on both concentration and polarization of the source (Fig.~\ref{fig:enhancements}c and d, respectively). This relationship allows us to boost enhancement by increasing source molar polarization, either through improved pulse sequences with better distant dipolar field suppression or by increasing the concentration of the DMM (Fig.~\ref{fig:f80}a,b). Both of these strategies were probed on our benchtop NMR setup and will be discussed below.

To ensure broad applicability and to prevent potential catalyst deactivation, which can occur with certain target moieties, we implemented an injection protocol for targets incompatible with the catalyst. For targets compatible with the catalyst, the injection step can be omitted. This avoids sample dilution and polarization loss due to T$_1$ relaxation of the source magnetization during target injection ($\approx$ 10 s total injection time). The typical source molar polarization in our experiments without injection lies in the range of 500--600 mM, such that omitting the injection step can in principle increase the achievable enhancement by a factor of about 1.8--2.2.

\subsection*{Predictive framework for PHIPNOE enhancements}

Having demonstrated the robustness, repeatability and versatility of the PHIPNOE approach, we now discuss the physical principles governing which molecules are suitable targets and how enhancements can be predicted. We developed a theoretical framework to predict target molecule PHIPNOE enhancements based on the Solomon equations for cross-relaxation. The key insight is that hyperpolarized source molecules transfer magnetization to target molecules through dipolar coupling, with the transfer efficiency depending on three main factors: the relaxation times of both species (T$_{1,\mathrm{source}}$ and T$_{1,\mathrm{target}}$), their molecular masses, and the initial source polarization.

In our simplest model (Model 1), we assume that the cross-relaxation rates (W$_0$ and W$_2$) are molecule-independent, allowing us to predict enhancements using only the T$_1$ values and initial source polarization. Model 1 captures the essential physics: enhancement is maximized when the target T$_1$ significantly exceeds the source T$_1$, which allows efficient polarization transfer before the source polarization decays. For molecules with similar T$_1$ values, the enhancement approaches a limiting value determined by the cross-relaxation efficiency.

\begin{figure}[h]
    \centering
    \includegraphics[width=\textwidth]{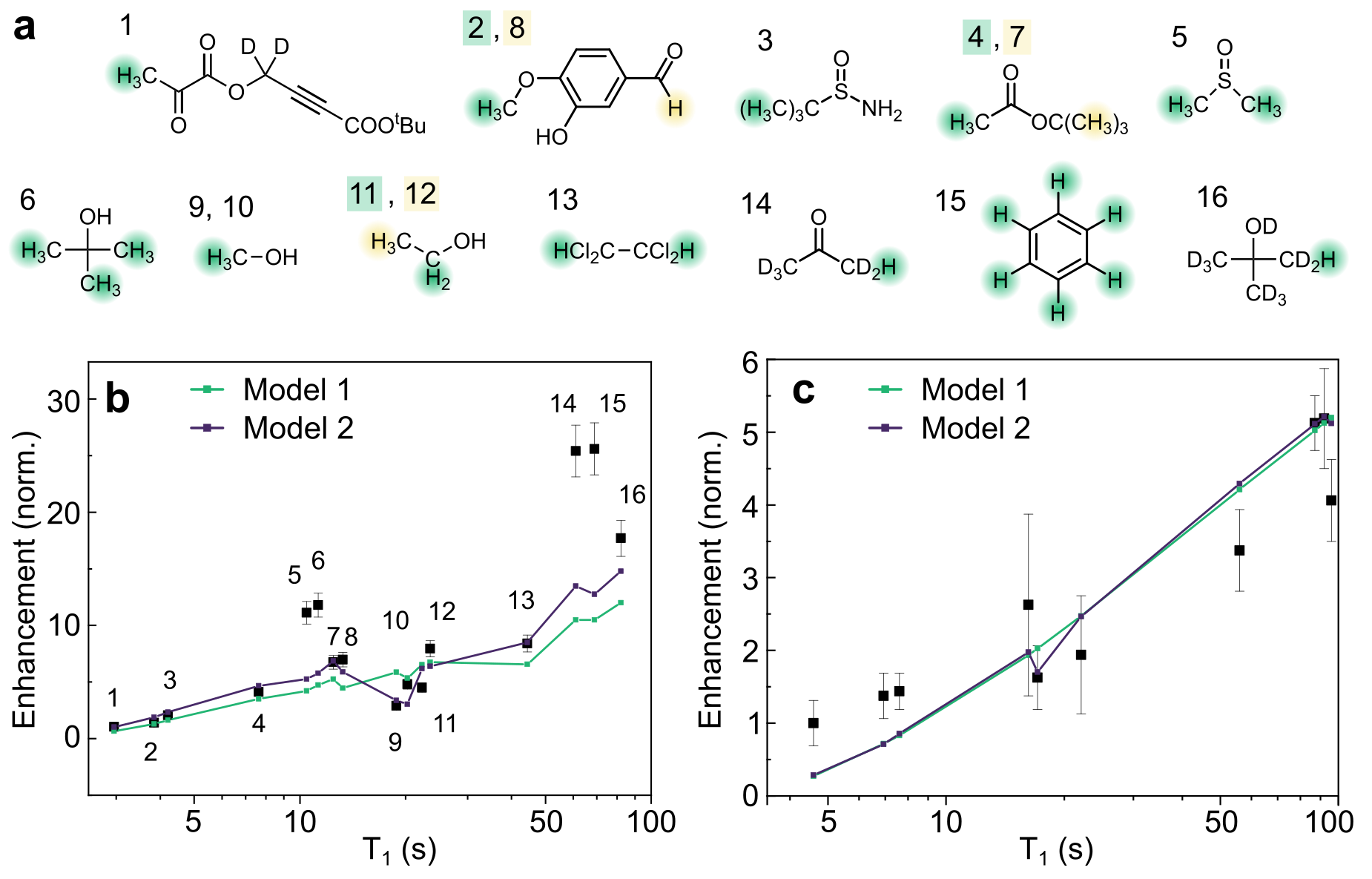}
    \caption{\textbf{Predictive models for PHIPNOE enhancement.} \textbf{a}, The molecular targets used to develop the predictive model. \textbf{b}, PHIPNOE enhancement on various protons in molecular targets, plotted against the relaxation time of the proton site. Key: (1) t-butyl propargyl-pyruvate ester\cite{nagel2023parahydrogen} (CH$_3$); (2) isovanillin (CH$_3$); (3) t-butyl sulfinamide (CH$_3$); (4) t-butyl acetate (methyl CH$_3$); (5) DMSO (CH$_3$); (6) t-butanol (CH$_3$); (7) t-butyl acetate (t-butyl CH$_3$); (8) isovanillin (CHO); (9) methanol (CH$_3$); (10) methanol (CH$_3$); (11) ethanol (CH$_2$); (12) ethanol (CH$_3$); (13) tetrachloroethane (CH); (14) acetone-d5 (CD$_2$H); (15) benzene (CH); (16) t-butanol-d9 (CD$_2$H). The calculated correlation coefficient for this data is 0.86, indicating strong positive correlation. The two lines represent predictions of the experimental results using Model 1 ($T_1$ only) and Model 2 ($T_1$ and molecular mass dependent). \textbf{c}, Our models are fitted to the normalized enhancement data from Salnikov \textit{et al.}~\cite{Salnikov2022}.}\label{fig:PHIPNOE-enhancements}
\end{figure}

A refined model (Model 2) incorporates molecular mass through its effect on diffusion and molecular encounters. Smaller molecules diffuse faster and approach each other more closely, enhancing the dipolar coupling that drives polarization transfer. Combining T$_1$ and mass dependencies improves accuracy, capturing the general trend that longer target T$_1$ values correlate with higher enhancements. The refined model incorporating molecular mass (violet line, Fig.~\ref{fig:PHIPNOE-enhancements}b and c) shows closer agreement across all targets, reducing the RMS deviation from 1.8 to 1.25. Notably, both models also fit previously published data from Salnikov \textit{et al.}~\cite{Salnikov2022} (Fig.~\ref{fig:PHIPNOE-enhancements}c), suggesting that the framework applies beyond our specific experimental conditions. While much more comprehensive microscopic theories of intermolecular cross-relaxation exist~\cite{halleCrossrelaxationMacromolecularSolvent2003a}, our simplified approach effectively captures the experimental trends across different datasets. These results confirm that PHIPNOE is governed by well-understood physical principles that enable rational optimization of experimental conditions.

Validating these models required precise measurement of enhancement factors under the complex spectral conditions (e.g., heavy radiation damping) created by intense hyperpolarized source signals. Our fitting procedure~\cite{Eichhorn2022} uses multiple Lorentzians to deconvolute overlapping signals (Fig.~\ref{fig:RD_and_kinetics}a), accounting for different phase corrections between source and target resonances due to radiation-damping effects. The resulting kinetic profiles (Fig.~\ref{fig:RD_and_kinetics}b) reveal the characteristic time evolution of PHIPNOE signals, validating the competing relaxation processes underlying our theoretical framework.

\subsection*{Benchtop NMR detection of PFAS}
PHIPNOE could be transformative for benchtop NMR spectroscopy (typically $\leq$\,\SI{2}{\tesla}). A key limitation of benchtop NMR is low sensitivity compared with high-field NMR~\cite{galvan_successful_2023, castaing-cordier_chapter_2021}. As PHIPNOE polarization is independent of the strength of the detection field, the enhancement in low-field NMR will be substantially higher. We demonstrate that PHIPNOE implemented on a portable benchtop NMR systems may provide a viable route for detecting PFAS in drinking water, by boosting the \textsuperscript{19}F NMR signals of these small molecules. As a proof-of-principle demonstration, we show PHIPNOE enhancement of the \textsuperscript{19}F NMR signals from two relevant PFAS, trifluoroacetic acid (TFA) and perfluorooctanoic acid (PFOA). For these experiments we implemented a shuttling system equipped with a magnetic shield on a Spinsolve 60 Ultra spectrometer capable of \textsuperscript{19}F detection ($B_0 = $\,\SI{1.41}{\tesla}; see the Methods and Supplementary Information sections for implementation details). In contrast to the \textsuperscript{13}C signal enhancement in benzene shown above, PFAS molecules lack protons (except for rapidly exchanging acidic ones) that can be exploited for INEPT assisted polarization transfer. However, due to the relatively high gyromagnetic ratio of \textsuperscript{19}F, direct NOE transfer was efficient enough to yield a signal enhancement of 55 for the CF\textsubscript{3} group of TFA, and enhancements of 150 and 86 for the CF\textsubscript{3} and CF\textsubscript{2} groups of PFOA, respectively. The hyperpolarized spectra were obtained with a hard 90$^{\circ}$ pulse, demonstrating that PHIPNOE of \textsuperscript{19}F spins does not require source signal suppression, unlike proton-based detection.

\begin{figure}[h]
    \centering
    \includegraphics[width=\textwidth]{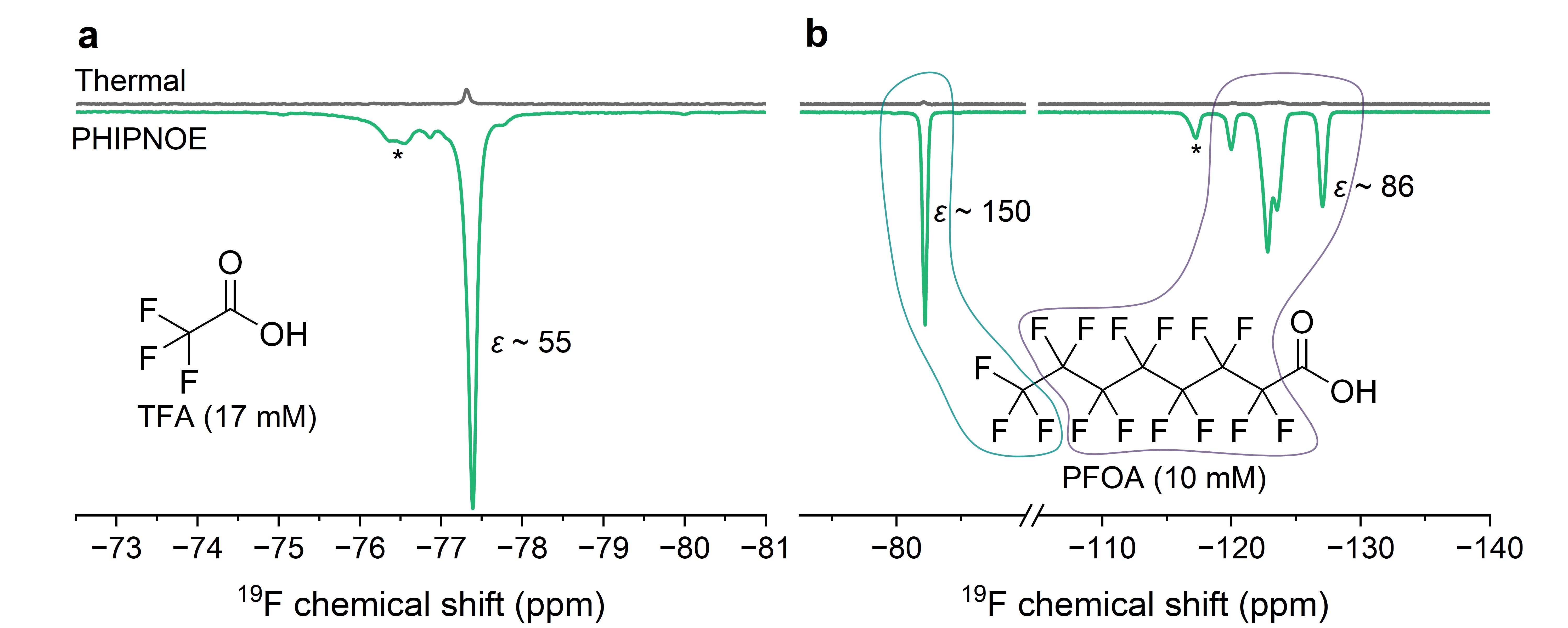}
    \caption{\textbf{PHIPNOE of \textsuperscript{19}F spins in PFAS molecules.} PHIPNOE spectra of TFA (\textbf{a}) and PFOA (\textbf{b}) obtained on a benchtop spectrometer. The asterisks indicate hyperpolarized signals that cannot be assigned to the free form of the PFAS under study, which most likely indicate interactions between the PFAS and the PHIP catalyst.}\label{fig:PIFAS}
\end{figure}

\subsection*{Maximizing enhancement at low field}

Having demonstrated PFAS detection on the benchtop, we explored ways to maximize PHIPNOE enhancements on a low-field system. For these experiments we used a second benchtop setup, a Bruker F80 benchtop spectrometer ($B_0 = $\,\SI{1.88}{T}; see the Methods and Supplementary Information sections), again equipped with a shuttling system and magnetic shield.

\begin{figure}[h]
    \centering
    \includegraphics[width=\textwidth]{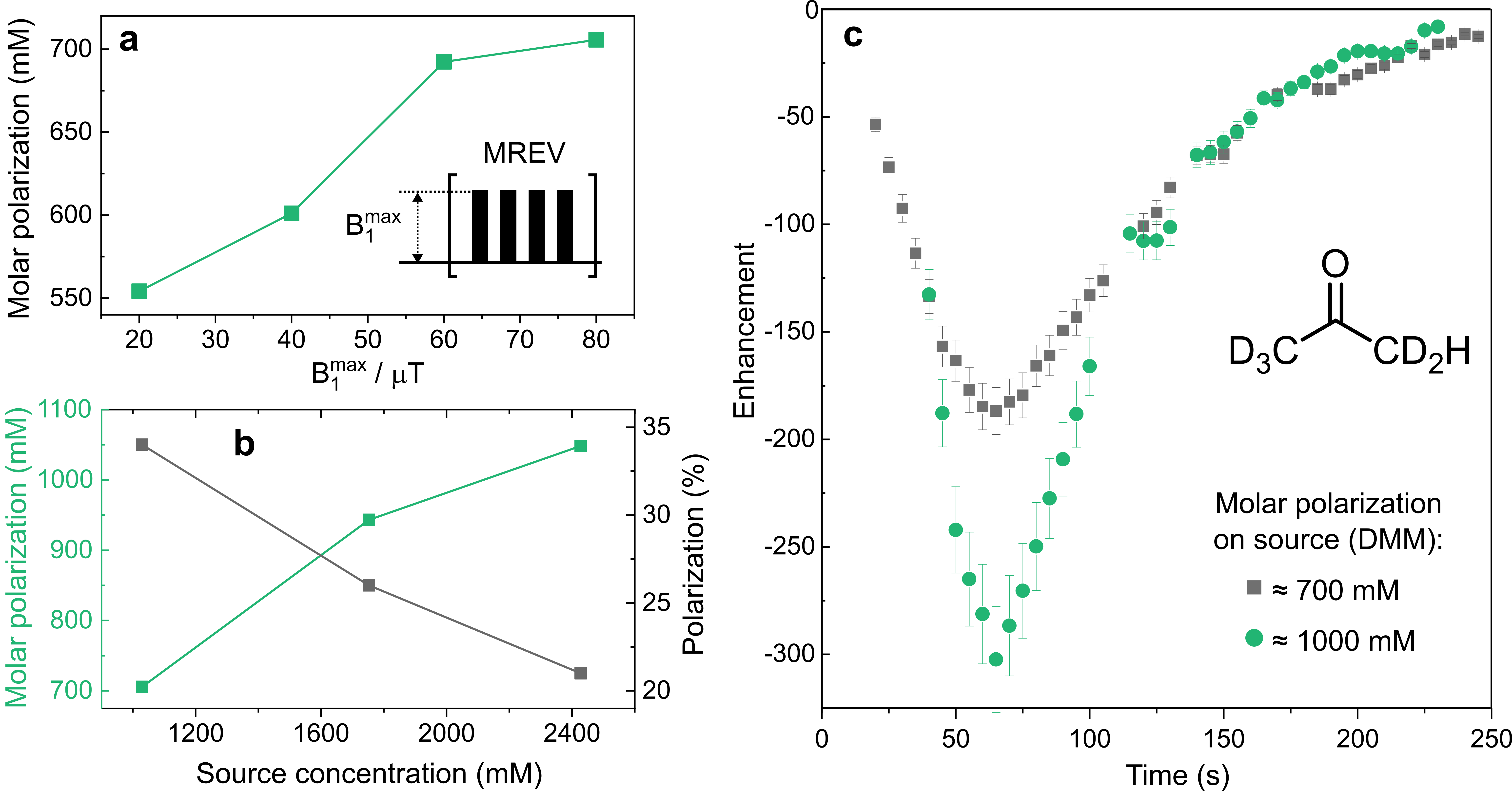}
    \caption{\textbf{PHIPNOE performance on a benchtop spectrometer.} \textbf{a}, The dependence of source spin molar polarization on the $B_1$ pulse amplitude in the MREV sequence. \textbf{b}, The dependence of source spin molar polarization and polarization on the concentration of the source molecule. \textbf{c}, The kinetics of PHIPNOE enhancement of $^1$H nuclei in acetone-d$_5$, for $\approx$\SI{700}{mM} and \SI{1000}{mM} molar polarization of the source species.}\label{fig:f80}
\end{figure}

The molar polarization of the source molecule dimethyl maleate for the 9.4~T data from Fig.~\ref{fig:PHIPNOE-enhancements} was \SI{280}{mM}. To improve PHIPNOE we took steps to increase the molar polarization.
To avoid diluting the source spins, we omitted the injection of a target molecule solution and observed PHIPNOE on the acetone-d\textsubscript{5} impurities in the solvent; combined with small improvements to the MREV sequence (see Methods for details), this improved source molar polarization from \SI{280}{mM} to $\approx$\SI{550}{mM}.
Next, we increased the amplitude of the $B_1$ pulses in the MREV sequence and scaled the number of pulses proportionally to $B_1$ to achieve better distant dipolar field decoupling (Fig.~\ref{fig:f80}a), reaching $\approx$\SI{700}{mM} molar polarization.
With these factors, we achieved a PHIPNOE enhancement of $\varepsilon \approx 185 \pm 11$ on the acetone isotopomer (Fig.~\ref{fig:f80}c).
This enhancement is over seven times higher than observed at high-field ($\varepsilon \approx 25$ at \SI{280}{mM}), but a linear extrapolation accounting for field strength and source molar polarization predicts $\varepsilon \approx 317$. We do not currently understand this difference, although it may be related to small differences in $T_1$ relaxation times of the molecules.

Finally, we investigated the limits of the technique by maximizing the initial substrate concentration from 1~M to 2.4~M to increase the total molar polarization, even at the cost of slightly reduced polarization percentage~\cite{Dagys2024} (Fig.~\ref{fig:f80}b). Under these conditions, we increased the source molar polarization from $\approx$\SI{700}{mM} to more than \SI{1}{M}, and the enhancement reached $300 \pm 25$. Although this is still below the predicted value of 450, a 300-fold signal amplification on a benchtop device confirms the utility of PHIPNOE for low-field analysis.

\subsection*{Discussion and outlook}

Our results establish PHIPNOE as a versatile platform for molecular hyperpolarization that operates under ambient conditions, without cryogenics or complex hardware adjustments. This makes it practical for routine NMR setups, including compact, benchtop spectrometers. Here we discuss the scope of the method, potential improvements and applications where PHIPNOE could have transformative impact.

\textbf{Target molecules.} In this work, we focused on the polarization of small target molecules with molecular weights ranging from 32 to 230 Da. Intermediate-sized molecules will typically display faster relaxation dynamics and PHIPNOE will likely be less effective for this class of molecules. However, hyperpolarization for these molecules is likely to be less beneficial as their short relaxation times enable fast signal averaging. For macromolecules, the relaxation dynamics again are favourable for polarization by PHIPNOE, based on hard-sphere simulations (see the Supplementary Information section). In addition, PHIPNOE may lead to high enhancements for surfaces in contact with the solution due to larger contact time for adsorbed molecules, similar to that observed for SPINOE with hyperpolarized xenon~\cite{room_enhancement_1997, pietrassSurfaceSelective1H1998}. Approaches for applications involving solvents other than acetone---and in particular aqueous solutions---are discussed in the following.

\textbf{Extending capabilities.} Three key areas present opportunities for extending PHIPNOE capabilities: expanding solvent compatibility, increasing achievable polarization levels and optimizing detection strategies.

Currently, we generate hyperpolarized DMM in acetone, which limits direct applications in aqueous or other solvent systems. Although we have shown that polarization transfer occurs upon mixing different solvents, a preferable solution is to use alternative PHIP substrates that can be directly polarized in the target solvent. For instance, parahydrogen-polarized fumarate can be produced in aqueous media at high concentrations~\cite{gierse2023parahydrogen}, and the resulting olefinic $^1$H spins have favourable relaxation properties for NOE transfer~\cite{ripka2018hyperpolarized}.

Target signals should increase with higher source polarization, which can be achieved through higher concentrations. Our data demonstrate a linear relationship between source and target polarization (Fig.~\ref{fig:enhancements}). By implementing stronger static fields and radiofrequency fields during the polarization transfer sequence, we project two- to three-fold improvements in source polarization. This would enable routine 50-fold enhancements at \SI{400}{\mega\hertz} for many small molecules.

Finally, the detection of the large signal from hyperpolarized source molecules can be further improved. Selective excitation sequences can be used to observe target molecules while preserving source magnetization for continued polarization transfer~\cite{hilty_hyperpolarized_2022}. Alternatively, solvent suppression techniques analogous to water suppression in biomolecular NMR can eliminate the source signal entirely~\cite{piotto_gradient-tailored_1992}. These approaches would be particularly valuable for mixture analysis or when target and source signals overlap. We demonstrate two approaches that address detection challenges in PHIPNOE experiments. Strong solvent suppression using WATERGATE-like sequences~\cite{piotto_gradient-tailored_1992} effectively eliminates the dominant DMM signal while preserving enhanced target resonances, enabling direct observation without interference from the intense source signal (Fig.~\ref{fig:teaser}b). Alternatively, PHIPNOE polarization can be transferred to heteronuclei through established pulse sequences: $^{13}$C detection of benzene using INEPT transfer provided a 31-fold enhancement relative to thermal $^{13}$C signals (Fig.~\ref{fig:teaser}c). Heteronuclear detection extends the scope of PHIPNOE beyond proton detection and provides the versatility needed for comprehensive molecular characterization.

\textbf{Applications.} 
PFAS are a growing concern as persistent 'forever chemicals' in drinking water.\cite{Hu2016} Targeted LC-MS/MS detects only a small fraction of the total organofluorine in a sample,\cite{Faber2025} whereas \textsuperscript{19}F NMR gives an inclusive, standard-free readout but is limited by its low sensitivity.\cite{Camdzic2023} PHIPNOE might address this limitation by improving the sensitivity of \textsuperscript{19}F NMR. Unlike costly HPLC-MS or conventional high-field NMR, PHIPNOE combined with benchtop NMR detection could provide a cost-effective, cryogen-free way to distinguish PFAS and their breakdown products with chemical specificity for real-time in-field applications.\cite{lingPFASQuantificationUsing2026}

With the demonstrated 30-fold enhancements at \SI{400}{\mega\hertz} and repeatability below 15\% standard deviation, PHIPNOE could directly benefit high-field NMR applications where measurement time limits throughput. This includes ligand binding screening for drug discovery~\cite{hajduk_nmr-based_1999, lepre_theory_2004} and identification of metabolite mixtures~\cite{markley_future_2017}. Similarly, the combination of PHIPNOE with ultrafast 2D NMR~\cite{Parker2023} should enable single-shot 2D NMR with high sensitivity. This is notable given that hyperpolarization cannot be replenished. A further opportunity is combining PHIPNOE with pure-shift NMR~\cite{taylor_sabre-enhanced_2021}, enabling cleaner chemical shift information without the typical reduction in sensitivity associated with such experiments~\cite{zangger_pure_2015}.

Finally, several key NMR investigations require monitoring processes that occur over the seconds-to-minutes timescale, such as nucleation experiments~\cite{weber_assessing_2020} and reaction monitoring~\cite{stevanato_real-time_2023, semenova_reaction_2019}. These are typically challenging processes to observe as the limited time for acquiring the NMR signal does not allow signal averaging. Hyperpolarization is a natural solution to these issues, and the fast repetition time achievable with PHIPNOE enables the exploration of a large parameter regime of the studied process.

By offering enhanced sensitivity and fast cycle times, PHIPNOE could extend hyperpolarized NMR beyond specialist research laboratories to industrial quality control, teaching environments and field-deployed analysis.

\section*{Methods}\label{sec:methods}

\subsection*{Experimental setup}

\textbf{Parahydrogen generation and delivery.} Parahydrogen at >95\% enrichment was generated by passing hydrogen gas (>99.999\% purity) through a commercial parahydrogen generator (Advanced Research Systems, U.S.A.) operating at \SI{25}{\K}. The parahydrogen was collected in aluminium cylinders, and used in experiments at a cylinder pressure of approximately \SI{20}{bar}.

\textbf{Sample handling system.} PHIP experiments were performed using pressurizable 5\,mm (outer diameter) NMR tubes to hold the sample and facilitate parahydrogen bubbling. Custom screw caps with an O-ring were used to seal the tube, and these had three ports to allow polytetrafluoroethylene (PTFE) capillaries (1/16\,in. outer diameter, 1/32\,in inner diameter) to pass into the NMR tube. The first capillary terminated at the bottom of the NMR tube and was used for bubbling gas through the samples. The second capillary terminated at the top of the NMR tube and served as a gas outlet. The third capillary terminated halfway down the NMR tube, and was used to inject solutions containing target molecules. Gas-flow control for PHIP experiments comprised mainly computer-controlled solenoid valves and stainless steel fluidic components, connected with 1/4" stainless steel tubing.

\textbf{Low-field polarization hardware.} For applying NMR pulse sequences at low field, a magnetic shield (MS-1L, Twinleaf LLC, USA) was used to provide a 10$^4$ shielding factor from the laboratory field. The residual field in the shield was on the order of nanotesla, as verified with a fluxgate magnetometer. Oscillatory and pulsed magnetic fields were produced using built-in B$_x$ and B$_y$ coils on a flexible PCB, powered with a multifunction analog/digital in/out card (NI-6343, National Instruments, USA) connected to 12\,W audio amplifiers (M032S, Kemo Electronic GmbH, Germany). The static \textit{z}-field was produced by a solenoid passing through the shield, powered by a programmable power supply (HMP2030, Rohde \& Schwarz, Germany). \textit{For the benchtop setup}, the home-built shield was constructed. Control of the static magnetic field B0 as well as the oscillating field B1 was achieved by using the analog output of an I/O card (BNC-2110, National Instruments, USA), which was subsequently amplified by home-built amplifiers. B0 driver is an Improved Howland Current Source using TI OPA548 and Analog Devices LT2057 and the B1 driver is a class AB current source with output power of 39\,dBm.

\textbf{Sample shuttling.} The sample was shuttled between this low-field region and the centre of a 9.4-T NMR magnet using a motorized mechanical shuttle system. Shuttle time from the low-field region to the sample position in the NMR spectrometer was typically on the order of 1 second. The mechanical shuttle comprises a linear stage powered by a stepper motor with a carbon-fibre rod attached to a 3D-printed sample holder.

\subsection*{Experimental protocols}

For all experiments, unless otherwise specified, the following experimental protocols were followed.

\textbf{Pulse sequence.} The modified MREV pulse sequence, used here to convert proton singlet order into observable magnetization, is given by the scheme:
\begin{equation*}
\left[\frac{\tau}{2} - 90_X - 2\tau - 90_X - \tau - 90_Y - 2\tau - 90_{\bar{Y}} - \tau \right.
\left. - 90_{\bar{X} + \phi} - 2\tau - 90_{\bar{X} + \phi} - \tau - 90_{Y + \phi} - 2\tau - 90_{\bar{Y} + \phi} - \frac{\tau}{2} \right]^N.
\end{equation*}

The resonance condition for the sequence is $\phi = 12 \pi J_{HH} \tau$, where $J_{HH}$ is the scalar coupling between the protons, which is \SI{13.2}{Hz} in our case. The polarization transfer is driven by the difference in the J-coupling constants of the two protons to carbon-13. In our case, this difference is $dJ = \left| J_{H_1C} - J_{H_2C}\right| = \left|13.2 - 2.1\right| = 11.1$ Hz. The pulse sequence scales this value down to an effective coupling of $dJ^* \approx (0.47+0.1\frac{\tau_{90}}{\tau})\, dJ$. To calculate the number of cycles required for the full transfer, the formula $n=\left[ \frac{1}{(dJ^*/\sqrt{2}) 12 \tau} \right]$ can be used. In our experiments, these values were further optimised experimentally. The best dipolar field suppression is achieved with the shortest $\tau$ possible, which allows fitting the largest number of decoupling cycles into the pulse sequence, which is limited by the length of a 90-degree pulse, so $\tau$ is slightly larger than $\tau_{90}$. Our experimental parameters were: $B_0=\SI{100}{\micro\tesla}$, $B_1= \SI{420}{Hz}$, $\tau=\SI{0.5958}{ms}$, and $n=32$ cycles. This results in a duration of the overall pulse sequence of \SI{0.228}{s}. \textit{For benchtop experiments}, Gaussian pulses with a 5\% truncation instead of square pulses were used to eliminate cross-talk. In addition, $B_0=\SI{380}{\micro\tesla}$ is larger than in high-field experiments, to eliminate cross-talk when driving larger $B_1$, the maximum amplitudes of which are shown in Fig.~\ref{fig:f80}a. The pulse sequence also lacked the phase shift on $Y$ pulses, while all other spin dynamics remained the same. Additional information on MREV calibration on the benchtop setup can be found in the~\hyperref[SI:MREV]{SI}. A detailed description of the theory and design principles for this class of pulse sequences is provided elsewhere~\cite{korzeczek_phip_2025, Korzetal24}.

\textbf{PHIP reaction solution.} The reaction solution for PHIP experiments was 1\,M [1-$^{13}$C]dimethyl acetylene dicarboxylate-d\textsubscript{6} and 10\,mM rhodium catalyst ([1,4-bis(diphenylphosphino)butane](1,5-cyclooctadiene)rhodium(I) tetrafluoroborate) in acetone-d$_6$. The dimethyl acetylene was added to the catalyst--acetone solution a few minutes or less before starting the PHIP reaction, to minimize the time for undesirable catalyst side reactions to occur. \SI{400}{\micro\L} of this solution was introduced into a 5-mm NMR tube, which was sealed with a screw cap, and loaded into the sample shuttle system.

\textbf{Polarization of the source molecules.} To polarize the source molecules, parahydrogen was bubbled through the reaction solution for \SI{28}{s} at 10-bar pressure. During this time, the solenoid provided a field of \SI{100}{\micro\tesla} and $^1$H decoupling was employed to negate polarization losses from singlet--triplet mixing~\cite{kating_nuclear_1993}. Subsequently, a modified MREV pulse sequence was applied to convert the proton singlet order into magnetization and to produce $^1$H dipolar decoupling to reduce polarization losses due to high internal sample magnetization. \textit{For the benchtop experiments}, the typical bubbling time was \SI{30}{s}. However, to maximize molar polarization (Fig.~\ref{fig:f80}b), higher concentrations were prepared using a prolonged bubbling duration of \SI{45}{s}. Specifically, a source concentration of $\approx$\SI{1700}{mM} was achieved by mixing \SI{440}{\micro\L} of \SI{2}{M} DMAD with \SI{120}{\micro\L} of \SI{15}{mM} catalyst solution. For the $\approx$\SI{2400}{mM} source concentration, \SI{380}{\micro\L} of \SI{2}{M} DMAD was mixed with \SI{120}{\micro\L} of \SI{15}{mM} catalyst solution. The higher final concentration relative to the initial mixture is attributed to rapid solvent evaporation during the prolonged gas bubbling phase.

\textbf{Data acquisition.} The sample was shuttled into the 9.4-T NMR system and a single-scan $^1$H NMR spectrum was acquired using a 0.5$^\circ$ pulse to perturb the spins only weakly. After that, spectra were acquired continuously every \SI{5}{s} using 5$^\circ$ flip-angle pulses, and \SI{200}{\micro\L} of target-molecule solution was injected using \SI{5}{bar} of N$_2$ pressure for \SI{8}{s}.

\textbf{Single-shot $^1$H spectra with solvent suppression.} To acquire a single-shot enhanced spectrum, shown in Fig.~\ref{fig:teaser}a, b, the solution of \textbf{a} (THF, ethanol, isopropanol and diethyl [difluoro(trimethylsilyl)methyl]phosphonate) or \textbf{b} (isovanillin molecule) was mixed with the initial catalyst stock solution, resulting in the final concentration of \SI{100}{mM}. There was no need for its injection after the PHIP procedure, since these molecules do not degrade the catalyst during the experiment. After transferring the sample back into the high-field spectrometer, a final DMM concentration of \SI{1000}{mM} and \SI{578}{mM} was measured with a polarization level of \SI{32}{\%} and \SI{27}{\%}, corresponding to a polarized concentration of \SI{604}{mM} and \SI{303}{mM}, respectively. The smaller molar polarization on iso-vanillin was achieved on purpose to get comparable data with the injection experiment, see Figure~\ref{fig:PHIPNOE-enhancements}. For the mixture, a 20-second waiting time after sample arrival was used. For isovanillin, a 10-second waiting time after sample arrival was used, as determined from a separate kinetic experiment. After this waiting period, a small-angle \ang{0.3} pulse was applied to acquire a spectrum in magnitude mode, allowing for the accurate determination of the DMM frequency. Subsequently, strong solvent suppression was performed using a sequence of 1000 Hz selective G4 pulses, followed by the application of additional \ang{90} pulses on the $^{13}$C channel and a strong gradient. This sequence was repeated 32 times. Finally, detection was performed with a WATERGATE suppression sequence. The resulting spectrum is displayed in Fig.~\ref{fig:teaser}a, b. A residual suppression artifact is likely due to the presence of a thick capillary within the NMR tube.

\textbf{INEPT PHIPNOE experiment.} For the INEPT sequence, the coupling was approximately $J_{CH} \approx 168$ Hz. The best INEPT parameters were determined on the thermal sample, with the first delay being 1.7 ms and the second 1.5 ms. The enhancement was approximately 2.4, which reduced transfer efficiency. This may be explained by the presence of a thick capillary to conduct PHIPNOE experiments. The T$_1$\{$^1$H\} on ($^{13}$C)C$_5$H$_6$ isotopomer was 28 s (\SI{400}{\mega\hertz}), which is approximately half that of the C$_6$H$_6$ benzene isotopomer, which had T$_1$\{$^1$H\} = 52 s. The sample volume was approximately 400 $\mu$L with 1M of DMAD and approximately 120 mM of benzene in acetone-d$_6$. The polarization procedure was identical to the one described before. For the INEPT experiment, the final concentration of DMM was approximately 900 mM with 33.15\% $^1$H polarization (0.59 M $^1$H molar polarization), and for the cross-relaxation experiment, the final concentration of DMM was 877 mM with 35.03\% $^1$H polarization (0.61 M $^1$H molar polarization). The cross-relaxation time after polarization transfer was set to be 28 s. The overall enhancements on $^{13}$C nuclei due to the relaxation alone was 20.4 compared to the thermal signal, whereas the enhancement due to the INEPT transfer was 31 compared to the INEPT signal, obtained for the thermal spectrum. The obtained signals are shown in Fig.~\ref{fig:teaser}c.

\bmhead{Supplementary information}

Supplementary information is available for this paper.

\bmhead{Acknowledgements}

This project was supported by the Initiative and Networking Fund of the Helmholtz Association (Project No. VH-NG-20-20), the ERC Synergy Grant HyperQ (Grant no. 856432), Research Council of Lithuania (Grant no. S-MIP-25-24), the BMFTR via projects QuE-MRT (grant no 13N16447), QMED2-PHIP-NMR (Grant no 03ZU2110CB), the European Innovation Council (EIC) project MagSense (Grant no. 101113079), NMR quantum sensing for environmental analysis (NQUA) (Grant no 13N16994). We thank Andreas Trabesinger for contributions to the preparation of this manuscript.

\section*{Declarations}

\textbf{Funding} See Acknowledgements.

\textbf{Conflict of interest} B.A.R., A.J.P., L.D., S.M., S.K., R.S., J.E. and I.S. are or were employees of NVision Imaging Technologies GmbH. I.S. and M.B.P. are cofounders and shareholders of NVision Imaging Technologies GmbH. NVision Imaging Technologies GmbH is working to commercialize products related to the reported research. All other authors declare that they have no competing interests.

\textbf{Author contributions} I.S., M.B.P., and S.K. initiated and supervised the research. A.J.P. and L.D. performed error estimation and investigated dependencies on target concentration, source polarization, and source concentration. J.E. conducted molecular screening experiments. M.K. developed the model to fit the data. R.S. engineered the benchtop setup. B.R. and S.M. performed demonstrator $^1$H experiments with signal suppression. F.D.B., R.W., P.M., S.M., and L.E. developed a solvent suppression sequence. K.V. conducted experiments with PFAS molecules on the benchtop, B.R. conducted all the remaining benchtop experiments and calculated proton enhancements using fitting procedures for all proton experiments. B.R., A.J.P., J.E., and I.S. wrote the manuscript with input from all authors.

\textbf{Data availability} The data supporting the findings of this study will be made publicly available upon publication of the peer-reviewed version of this work.

\appendix

\section{Estimating PHIPNOE enhancements}\label{secA1}

Using $\mathrm{s}$ as source and $\mathrm{t}$ as target with their respective magnetizations $M_\mathrm{s}(t), M_\mathrm{t}(t)$
we can use the Solomon equations from \cite{solomonRelaxationProcessesSystem1955} (We adapt their equations (9) to our notation and here differentiate between the two T$_1$ times $T_{\mathrm{1,s}}$ and $T_{\mathrm{1,t}}$. In the reference, these are the same but we assume some non-zero intrinsic T$_1$ times for the two spin types, which are different in general.) and the hyperpolarized source assumption $(M_\mathrm{s}(0),M_\mathrm{t}(0))=(M_\mathrm{s},M_\mathrm{t,th})$ and $M_\mathrm{s}(t)\gg M_\mathrm{t}(t),M_\mathrm{s,th}(t)$ to reach an approximate dynamics:

\begin{align*}
\dfrac{d}{dt} \ M_\mathrm{s}(t) &= - \dfrac{1}{T_\mathrm{1,s}} (M_\mathrm{s}(t)-\underset{\approx 0}{\underbrace{M_\mathrm{s,th}}}) + \underset{\approx 0}{\underbrace{(W_0 - W_2) M_\mathrm{t}(t)}} \\
\dfrac{d}{dt} \ M_\mathrm{t}(t) &= - \dfrac{1}{T_\mathrm{1,t}} (M_\mathrm{t}(t)-M_\mathrm{t,th}) + (W_0 - W_2) M_\mathrm{s}(t)
\end{align*}

In these approximations $M_\mathrm{s}(t)= M_s e^{-t/T_\mathrm{1,s}}$ so that we are left with

\begin{align*}
\dfrac{d}{d t} \ M_\mathrm{t}(t) &= - \dfrac{1}{T_\mathrm{1,t}} (M_\mathrm{t}(t)-M_\mathrm{t,th}) + (W_0 - W_2) M_s e^{-t/T_\mathrm{1,s}}
\end{align*}
which can be directly integrated (assuming $T_\mathrm{1,s}\neq T_\mathrm{1,t}$) to give
\begin{align*}
    M_\mathrm{t}(t) = M_\mathrm{t,th}+ M_s \dfrac{W_2-W_0}{1/T_\mathrm{1,s}-1/T_\mathrm{1,t}} (e^{-t/T_\mathrm{1,s}}-e^{-t/T_\mathrm{1,t}})
\end{align*}

which is extremized at
\begin{align*}
    \frac{d}{d t} M_\mathrm{t}(t^\ast) \overset{!}{=} 0 \Rightarrow t^\ast = \dfrac{\ln(T_\mathrm{1,s}/T_\mathrm{1,t})}{1/T_\mathrm{1,t}-1/T_\mathrm{1,s}},
\end{align*}
giving us
\begin{align*}
M_\mathrm{t}(t^\ast) = M_\mathrm{t,th}- M_s (W_2-W_0) (T_\mathrm{1,t})^{\frac{-T_\mathrm{1,s}}{T_\mathrm{1,t}-T_\mathrm{1,s}}}(T_\mathrm{1,s})^{\frac{T_\mathrm{1,t}}{T_\mathrm{1,t}-T_\mathrm{1,s}}}
\end{align*}

We therefore have an effective coupling strength $W_2-W_0$, which is multiplied by the initial (hyper)polarization $M_\mathrm{s}$ and an effective time that is smaller than either of $T_\mathrm{1,s}$ and $T_\mathrm{1,t}$ to give the change of the target polarization due to NOE.

The most straightforward prediction of experimentally reachable enhancement across molecules assumes that $W_2-W_0$ is independent of the molecule. This allows us to estimate $W_2-W_0$ as a single free parameter from the experiment and then make predictions using the above equation by inserting the correct $T_\mathrm{1,s},T_\mathrm{1,t}$ and initial $M_\mathrm{s}$.

We use a minimisation of the root mean square (RMS) difference between the experimentally measured enhancement and the model prediction across the molecules to determine the free rate parameter in the model. For this, the differences are normalised to the experimental error estimates. In Fig.~\ref{fig:PHIPNOE-enhancements}, the resulting predictions are shown as the curve `Model 1' with an RMS of $1.8$.

The second model additionally incorporates the molecular mass into an estimate for the rates $W_0$ and $W_2$:
\begin{align*}
W_{0}\propto {\frac {2\tau _{c}}{(1+(\omega _{s}-\omega _{t})^{2}\tau _{c}^{2})}}{\frac {1}{r^{6}}} \\
W_{2}\propto {\frac {12\tau _{c}}{(1+(\omega _{s}+\omega _{t})^{2}\tau _{c}^{2})}}{\frac {1}{r^{6}}}
\end{align*}

sharing the same proportionality factor. In our case, we have protons for both signal and target such that $\omega_\mathrm{s}\approx\omega_\mathrm{t}$. This yields
\begin{align*}
(W_2-W_0) \propto \dfrac{\tau_c}{r^6} \left(  \dfrac{12}{1+4\,\omega^2_\mathrm{s}\tau^2_c} -2 \right)
\end{align*}
where $r$ is the effective contact distance of the signal and target molecules and $\tau_c$ is the corresponding thermal time scale.

A simple estimate assumes kinetic diffusion in three-dimensional space which relates $\tau_c$ and $r$ with $r = \sqrt{6D\tau_c}$ where $D$ is the relevant diffusion constant. As an estimate for the mass dependence, we use the expected scaling from hard spheres in the kinetic theory of gases \cite{chapmanXIIKineticTheory1997} which follows $D\propto \sqrt{\dfrac{1}{m_\mathrm{s}}+\dfrac{1}{m_\mathrm{t}}}\ \dfrac{1}{(r_\mathrm{s}+r_\mathrm{t})^2}$ with the molecular masses $m_\mathrm{s},m_\mathrm{t}$ and the radii $r_\mathrm{s},r_\mathrm{t}$.

We approximate the molecules as following a bulk relationship with the same density such that $r = r_\mathrm{s}+r_\mathrm{t}\propto m_\mathrm{s}^{1/3}+m_\mathrm{t}^{1/3}$.

Combining these estimates, we reach the `Model 2' predictions in Fig.~\ref{fig:PHIPNOE-enhancements}. Here, two parameters are fitted to the data, namely a rate prefactor as in the `Model 2' case and an additional prefactor for $\tau_c$ which effectively describes how strongly the extreme narrowing regime of $\omega_\mathrm{s}\tau_c \ll 1$ is fulfilled. The Model 2 reaches an RMS deviation of $1.25$ compared to the experimental results.

\section{Data processing for enhancement evaluation}\label{secA2}

To determine the enhancement values, we reproduced the protocol described before \cite{Eichhorn2022}: the overall spectra (both real and imaginary parts) were fitted by several Lorentzians. One Lorentzian was specifically assigned to the strong signal originating from the dimethyl maleate molecule, while the others corresponded to the signals of interest. We note that the phase corrections for the source and the other signals could vary considerably. This is because the radiation damping (RD) from the intense source signal can act as a weak perturbing pulse on its surroundings. To avoid interference from irrelevant signals, the spectral regions used for fitting were selected manually.

\begin{figure}[h]
    \centering
    \includegraphics[width=\textwidth]{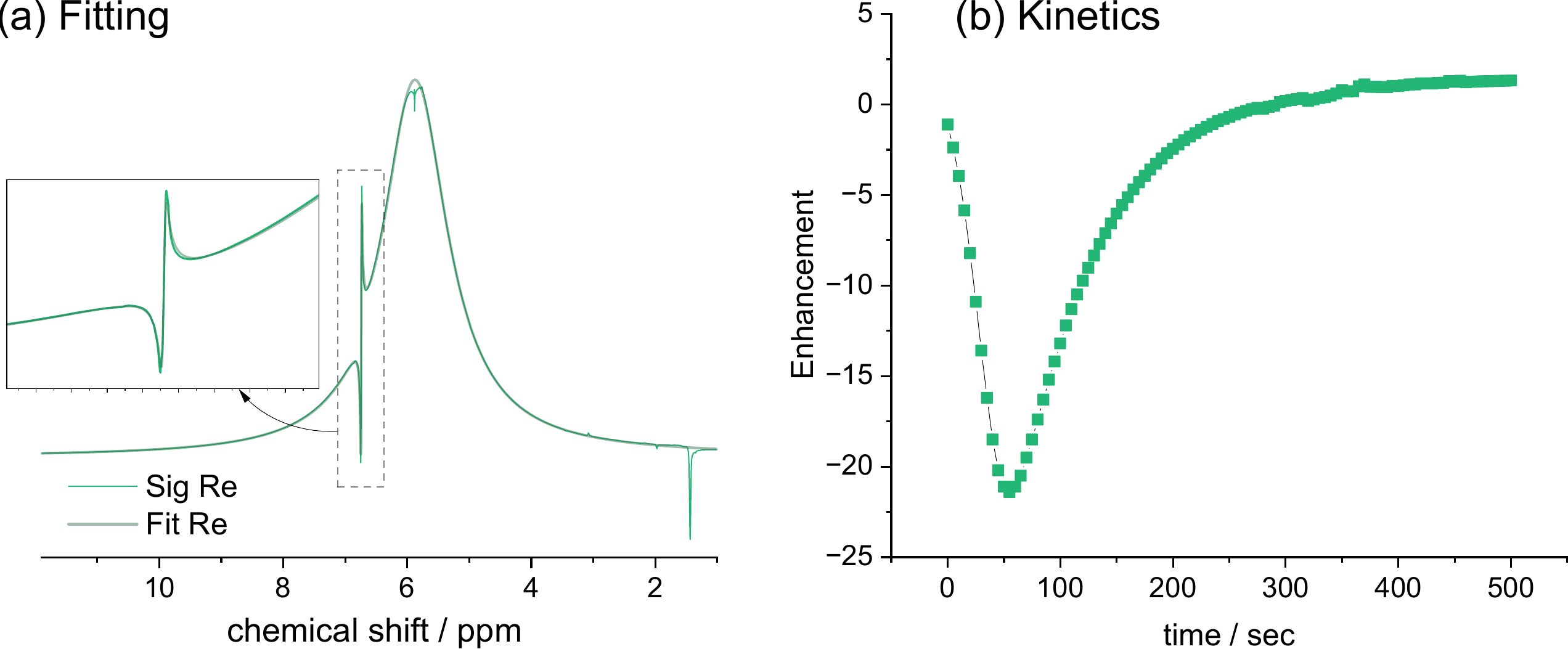}
    \caption{\textbf{Spectral analysis and kinetics of PHIPNOE transfer.} (a) Example of the fitting procedure used to extract enhancements for kinetic build-up and decay during the PHIPNOE process by using small flip angle excitation. (b) Representative kinetics for benzene molecules.}\label{fig:RD_and_kinetics}
\end{figure}

\section{Carrier frequency calibration for modified MREV pulse sequence on the benchtop setup}\label{SI:MREV}

\begin{figure}[h]
    \centering
    \includegraphics[width=\textwidth]{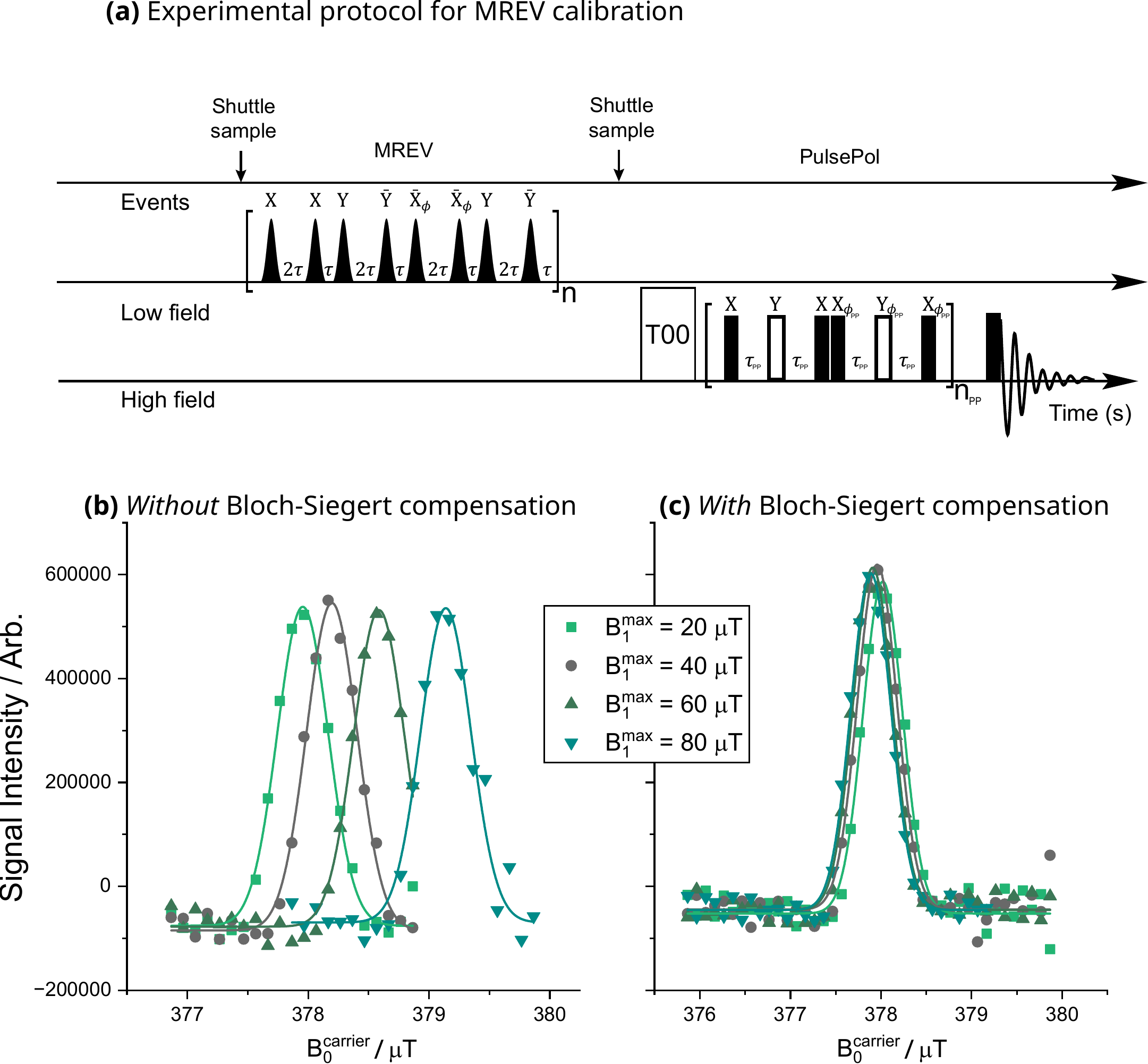}
    \caption{\textbf{MREV carrier frequency calibration for different pulse amplitudes.} \textbf{a}, The overall experimental protocol. Filled block boxes represent 90-degree pulses, while the open blocks represent 180-degree pulses. For the PulsePol sequence, the parameters were $\phi_{\text{PP}}=0.4\pi$, $4\tau_{\text{PP}}=\SI{20.5}{\milli s}$ and $n_{\text{PP}}=30$. The T00 filter consists of $\{G_1 - 90_{54.735}-G_2-90_{54.735}-90_{180}-G_3\}$, where $G_{i}$ means the sine gradient with \SI{2}{\milli s} recovery time, their durations are $d_{\{1, 2, 3\}} =\{4.4,\; 2.4, \; 2\}$~ms and their amplitudes are ${\{8.7,\; 5.1,\; 6.9\}}$~G/cm. \textbf{b}, The dependency of the signal after the experimental protocol described in \textbf{a} without Bloch--Siegert compensation. \textbf{c}, Same as \textbf{b}, but with Bloch--Siegert compensation.}\label{fig:B0_callibration}
\end{figure}

The MREV pulse sequence exhibits a narrow tolerance for field deviation, with a full width at half maximum (FWHM) efficiency bandwidth of only $\approx$ \SI{0.6}{\micro\tesla}. Consequently, it is critical to calibrate the carrier frequency of the applied pulses to strictly match the Larmor frequency of the spins. The overall experimental protocol for this is shown in Fig.~\ref{fig:B0_callibration}a. First, the sample is shuttled into the low-field region. There, the MREV sequence is applied with a carrier frequency set to $\omega^{\text{carrier}} = B_0^{\text{carrier}} \cdot \gamma_\text{H}$, where $\gamma_{\text{H}}$ is the proton gyromagnetic ratio. Subsequently, the sample is shuttled back to the high-field region. The resulting spin state is processed using a $T_{00}$ filter~\cite{pileioLonglivedNuclearSpin2020a} to isolate the singlet order, which is then converted into observable magnetization using a PulsePol sequence~\cite{schwartzRobustOpticalPolarization2018}. The singlet order destruction (SOD) filter was used to speed up the consecutive experiments~\cite{rodinFastDestructionSinglet2019}.

As shown in Fig.~\ref{fig:B0_callibration}b, increasing the maximal amplitude of the pulses in the MREV sequence results in a shift of the optimal $B_0^{\text{carrier}}$. This deviation arises from the Bloch--Siegert shift. This effect can be mitigated by applying Bloch--Siegert compensation, as described elsewhere~\cite{zeuchExactRotatingWave2020}. The result of this correction is shown in Fig.~\ref{fig:B0_callibration}c, where the optimal carrier frequency remains stable across different pulse amplitudes.

To adjust the MREV sequence parameters for different field strengths, the number of cycles and the inter-pulse delay $\tau$ were recalculated (see Fig.~\ref{fig:B0_callibration}a). The scaling factor for a Gaussian pulse with 5\% truncation is $s=1.98$. For a reference amplitude of $B_1^{\text{max}}=\SI{20}{\micro T}$, the sequence used $n_\text{MREV}=32$ cycles and $\tau=\SI{0.5958}{\milli s}$. For higher amplitudes scaled by a factor $k \in \{2, 3, 4\}$, the parameters were redefined as $B_1^{\text{max}} \rightarrow k \cdot B_1^{\text{max}}$, $n_{\text{MREV}} \rightarrow k \cdot n_{\text{MREV}}$, and $\tau \rightarrow \tau / k$.

\section{Phenomenological analysis of PHIPNOE kinetics}\label{secA4}

To characterize the time-dependent polarization dynamics, we monitored the signal enhancement $\epsilon(t)$ of the target resonances following the injection of the target solution into the hyperpolarized source solution. The observed kinetics are governed by two competing processes: the intermolecular NOE polarization transfer (build-up) and the spin-lattice relaxation (decay).

We modeled the experimental data using a phenomenological stretched bi-exponential function:

\begin{equation}
\epsilon(t) =
\begin{cases}
0 & t < t_{\text{start}} \\
A \cdot \left[ \exp\left(-\frac{(t - t_{\text{start}})^\beta}{\tau_{\text{build}}}\right) - \exp\left(-\frac{t - t_{\text{start}}}{\tau_{\text{decay}}}\right) \right] + C & t \ge t_{\text{start}}
\end{cases}
\label{eq:biexp_kinetics}
\end{equation}

where $\tau_{\text{build}}$ and $\tau_{\text{decay}}$ correspond to the characteristic time constants for polarization build-up and relaxation respectively; $A$ is an amplitude scaling factor; $C$ represents the thermal equilibrium baseline; $t_{\text{start}}$ accounts for the variable delay between the start of acquisition and the arrival of the hyperpolarized sample in the detection coil; and $\beta$ is a stretching exponent.

\textbf{Impact of injection dynamics.} We note that the build-up phase of the kinetics is convolved with the sample injection profile, which spans approximately 8 seconds. This non-instantaneous mixing distorts the early-time kinetics, deviating from the ideal instantaneous-mixing regime described by standard Solomon equations. To account for this, the stretching parameter $\beta$ was introduced to empirically model the broadened build-up profile.

We focused our primary analysis on the maximum achievable enhancement and the subsequent decay rates, which are robust to mixing dynamics. However, Equation~\ref{eq:biexp_kinetics} agrees well with the experimental data across the full time course. The complete dataset of fitted kinetic curves, along with the Python code used for non-linear least squares minimization (using the \texttt{lmfit} library~\cite{newvilleLMFITNonLinearLeastSquare2014}), is available in the associated data repository.

\begin{figure}[h]
    \centering
    \includegraphics[width=0.7\textwidth]{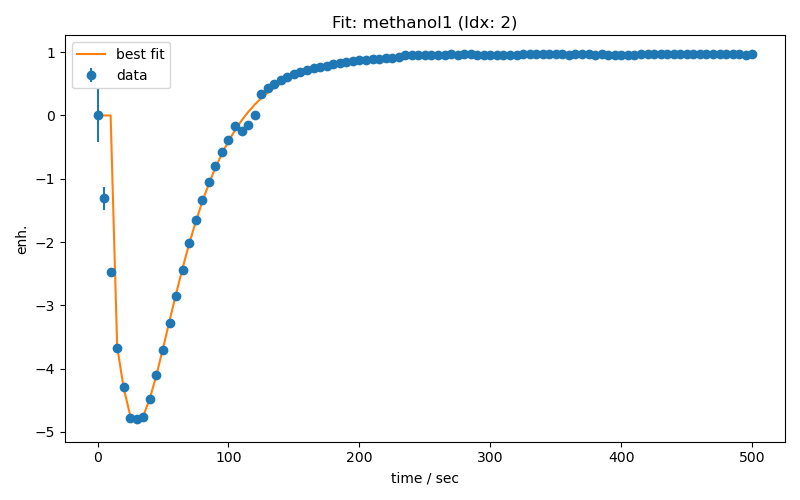}
    \caption{\textbf{Representative PHIPNOE kinetic profile.} The signal enhancement of methanol protons is plotted as a function of time. The solid orange line represents the best fit to the stretched bi-exponential model (Eq.~\ref{eq:biexp_kinetics}), capturing both the injection-limited build-up and the relaxation decay.}\label{fig:kinetics_model}
\end{figure}

\bibliography{sn-bibliography}

\end{document}